\documentclass[twocolumn]{article}
\usepackage{epsfig,amssymb}
\def\comment#1{}
\begin{document}
\begin{flushleft}{\Large{\bf{\textsf{Quantum-Enhanced Measurements: }}}}
\vskip .3\baselineskip
{\Large{\bf{\textsf{Beating the Standard Quantum
          Limit}}}}
\end{flushleft}
  
\begin{flushleft}{{{\bf{\textsf{Vittorio~Giovannetti$^1$, Seth~Lloyd$^{2}$,
            Lorenzo~Maccone$^3$}}}}\\{\scriptsize$^1$ {\it NEST-INFM
        \& Scuola Normale Superiore, Piazza dei Cavalieri 7, I-56126,
        Pisa, Italy.}}\\{\scriptsize$^2$ {\it MIT, Research Laboratory
        of Electronics and Dept. of Mechanical Engineering, 77
        Massachusetts Ave., Cambridge, MA 02139,
        USA.}}\\{\scriptsize$^3$ {\it QUIT - Quantum Information
        Theory Group, Dip. di Fisica ``A. Volta'', Universit\`a di
        Pavia, via A. Bassi 6 I-27100, Pavia, Italy.}}}
\end{flushleft}

\vbox{{\scriptsize\sf One sentence summary: {\bf To attain the limits
      to measurement precision imposed by quantum mechanics, `quantum
      tricks' are often required.}}}
\vskip 1\baselineskip
\setlength{\unitlength}{.22cm}
\begin{center}\begin{picture}(40,.01){\thicklines
\put(0,0){\line(1,0){40}}}\end{picture}
\end{center}
\begin{center}
  \parbox{8cm}{ {\sf Abstract:} Quantum mechanics, through the
    Heisenberg uncertainty principle, imposes limits to the precision
    of measurement. Conventional measurement techniques typically fail
    to reach these limits.  Conventional bounds to the precision of
    measurements such as the shot noise limit or the standard quantum
    limit are not as fundamental as the Heisenberg limits, and can be
    beaten using quantum strategies that employ `quantum tricks' such
    as squeezing and entanglement.  }
\end{center}
\vskip 1\baselineskip

\begin{figure}[h!]
\begin{center}
\epsfxsize=.81\hsize\leavevmode\epsffile{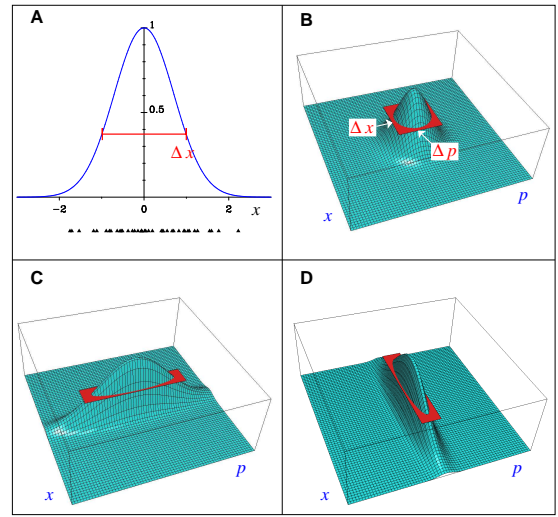}
\end{center}
\vspace{-.5cm}
\caption{The Heisenberg uncertainty relation. {\small{In quantum
      mechanics the outcomes $x_1$, $x_2$, {\em etc.} of the
      measurements of a physical quantity $x$ are statistical
      variables; that is, they are randomly distributed according to a
      probability determined by the state of the system. A measure of
      the ``sharpness'' of a measurement is given by the spread
      $\Delta x$ of the outcomes: An example is given in {\sf{\bf
          (A)}}, where the outcomes (tiny triangles) are distributed
      according to a Gaussian probability with standard deviation
      $\Delta x$. The Heisenberg uncertainty relation states that when
      simultaneously measuring incompatible observables such as
      position $x$ and momentum $p$ the product of the spreads is
      lower bounded: $\Delta x\:\Delta p\geq\hbar/2$, where $\hbar$ is
      the Planck constant.  The same is true when measuring one of the
      observables (say $x$) on a set of particles prepared with a
      spread $\Delta p$ on the other observable.  [In the general case
      when we are measuring two observables $A$ and $B$, the lower
      bound is given by the expectation value of the commutator
      between the quantum operators associated to $A$ and $B$.] In
      {\bf\sf (B)} we see a coherent state (depicted through its
      Wigner function): it has the same spreads in position and
      momentum $\Delta x=\Delta p$.  In {\sf\bf (C)} and {\sf\bf (D)},
      squeezed states are shown: they have reduced fluctuations in one
      of the two incompatible observables [i.e.  $x$ for {\sf\bf (C)}
      and $p$ for {\sf\bf (D)}] at the expense of increased
      fluctuations in the other.  The Heisenberg relation states that
      the red areas in the plots (given by the product $\Delta x\Delta
      p$) must have a surface larger than $\hbar/2$. In quantum
      optics, the observables $x$ and $p$ are replaced by the in-phase
      and out-of-phase amplitudes of the electromagnetic field, i.e.
      by its ``quadratures''. The Heisenberg principle is so called
      only for historical reasons: it is not a principle in modern
      quantum mechanics, since it is a consequence of the measurement
      postulate~\cite{robertson}.  Moreover, Heisenberg's formulation
      of a dynamical disturbance necessarily induced on a system by a
      measurement was experimentally proven
      wrong~\cite{scullywalther}: it is possible to devise experiments
      where the disturbance is totally negligible, but where the
      Heisenberg relations are still valid. They are enforced by the
      complementarity of quantum mechanics.}}}
\label{f:heisenb}\end{figure}

Measurement is a physical process, and the accuracy to which
measurements can be performed is governed by the laws of physics.  In
particular, the behavior of systems at small scales is governed by the
laws of quantum mechanics, which place limits on the accuracy to which
measurements can be performed.  These limits to accuracy take two
forms.  First, the Heisenberg uncertainty relation~\cite{robertson}
imposes an intrinsic uncertainty in the values of measurement results
of complementary observables such as position and momentum, or the
different components of the angular momentum of a rotating object
(Fig.~\ref{f:heisenb}).  Second, every measurement apparatus is itself
a quantum system: as a result, the uncertainty relations together with
other quantum constraints on the speed of evolution (such as the
Margolus-Levitin theorem~\cite{margolus}) impose limits on how
accurately we can measure quantities given the amount of physical
resources, e.g. energy, at hand to perform the measurement.

One important consequence of the physical nature of measurement is the
so-called `quantum back action': the extraction of information from a
system can give rise to a feedback effect in which the system
configuration after the measurement is determined by the measurement
outcome.  For example, the most extreme case (the so-called von
Neumann or projective measurement) produces a complete determination
of the post-measurement state.  When performing successive
measurements, quantum back action can be detrimental, as earlier
measurements can negatively influence successive ones. A common
strategy to get around the negative effect of back action and of
Heisenberg uncertainty is to design an experimental apparatus that
monitors only one out of a set of incompatible observables: `less is
more'~\cite{rmp}.  This strategy--- called quantum non-demolition
measurement~\cite{rmp,grangierqnd,wallqnd,brajinskij}--- is not as
simple as it sounds: one has to account for the system's interaction
with the external environment, which tends to extract and disperse
information, and for the system dynamics, which can combine the
measured observable with incompatible ones. Another strategy to get
around the Heisenberg uncertainty is to employ a quantum state in
which the uncertainty in the observable to be monitored is very small
(at the cost of a very large uncertainty in the complementary
observable). The research on quantum-enhanced measurements was spawned
by the invention of such techniques~\cite{rmp,cavesprd,yurke} and by
the birth of more rigorous treatments of quantum measurements~\cite{helstrom}.

Most standard measurement techniques do not account for these quantum
subtleties, so that their precision is limited by otherwise avoidable
sources of errors.  Typical examples are the environment induced noise
from vacuum fluctuations (the so-called shot noise) that affects the
measurement of the electromagnetic field amplitude, and the
dynamically-induced noise in the position measurement of a free mass
(the so-called standard quantum limit~\cite{sql}).  These sources of
imprecision are not as fundamental as the unavoidable Heisenberg
uncertainty relations, as they originate only from a non-optimal
choice of measurement strategy.  However, the shot-noise and standard
quantum limits set important benchmarks for the quality of a
measurement, and they provide an interesting challenge to devise
quantum strategies that can defeat them.  It is intriguing that almost
thirty years after its introduction~\cite{sql}, the standard quantum
limit has not yet been beaten experimentally in a repeated measurement
of a test mass.  In the meantime, a paradigm shift has occurred:
quantum mechanics, which used to be just the object of investigation,
is now viewed as a tool, a source of exotic and funky effects that can
be used to our benefit.  In measurement and elsewhere, we are
witnessing the birth of quantum technology.

We describe some of the techniques that have been recently developed
to overcome the limitations of classical measurement strategies.  We
start with a brief overview of some techniques to beat the shot noise
limit in interferometry.  In the process, we provide a simple example
that explains the idea behind many quantum-enhanced measurement
strategies.  We then give an overlook on some of the most promising
quantum-technology proposals and analyze the standard quantum limit on
repeated position measurements. Finally, we show the ultimate
resolution achievable in measuring time and space according to the
known physical laws. A caveat is in order: this review cannot be in
any way be viewed as complete because the improvement of
interferometry and measurements through non-classical light is at the
heart of modern quantum optics.  Many more ideas and experiments have
been devised than can be possibly reported here.

\begin{figure}[hbt]
\begin{center}
\epsfxsize=.8\hsize\leavevmode\epsffile{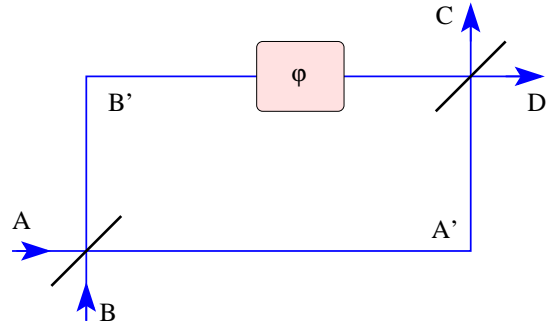}
\end{center}
\caption{Mach-Zehnder interferometer. The light field enters
  the apparatus through the input ports~A and~B of the first beam
  splitter and leaves it through the output ports~C and~D of the
  second beam splitter. By measuring the intensities (photon number
  per second) of the output beams one can recover the phase difference
  $\varphi$ between the two internal optical paths~A' and~B'.
  Formally, the input-output relation of the apparatus is completely
  characterized by assigning the transformations of the annihilation
  operators $a$, $b$, $c$ and $d$ associated to the fields at A, B, C
  and D, respectively. These are $c \equiv (a' + i e^{i\varphi} b')
  /\sqrt{2}$ and $d \equiv (i a' + e^{i\varphi} b') /\sqrt{2}$ with $
  a'\equiv (a + i b) /\sqrt{2}$ and $ b'\equiv (i a + b) /\sqrt{2}$
  the annihilation operators associated to the internal paths A' and
  B', respectively.}
\label{f:mzint}\end{figure}

\section*{\sf Interferometry: beating the shot noise limit}
In this section we focus on the issues arising in ultra-precise
interferometric measurements.  A prototypical apparatus is the
Mach-Zehnder interferometer (Fig.~\ref{f:mzint}). It acts in the
following way. A light beam impinges on a semi-transparent mirror
(i.e. a {beam splitter}), which divides it into a reflected and a
transmitted part.  These two components travel along different paths
and then are recombined by a second beam splitter. Information on the
phase difference $\varphi$ between the two optical paths of the
interferometer can be extracted by monitoring the two output beams,
typically by measuring their intensity (i.e. the photon number).  To
see how this works, suppose that a classical coherent beam with $N$
average photons enters the interferometer through the input~A. If
there is no phase difference $\varphi$, all the photons will exit the
apparatus at output~D.  On the other hand, if $\varphi=\pi$ radians,
all the photons will exit at output~C. In the intermediate situations,
a fraction $\cos^2(\varphi/2)$ of the photons will exit at the
output~D and a fraction $\sin^2(\varphi/2)$ at the output~C. By
measuring the intensity at the two output ports one can estimate the
value of $\varphi$ with a statistical error proportional to
$1/\sqrt{N}$. This is a consequence of the quantized nature of
electromagnetic field and of the Poissonian statistics of classical
light, which, in some sense, prevents any cooperative behavior among
the photons.  In fact, the quantity $\cos^2(\varphi/2)$ can be
experimentally obtained as the statistical average
$\sum_{j=1}^Nx_j/N$, where $x_j$ takes the value $0$ or $1$ depending
on whether the $j$th photon in the beam was detected at output~C or~D
respectively.  Because the $x_j$s are independent stochastic variables
(photons in the classical beam are uncorrelated), the variance
associated with their average is the average of the variances (central
limit theorem): the error associated with the measurement of
$\cos^2(\varphi/2)$ is given by
$\Delta(\sum_{j=1}^Nx_j/N)\equiv\sqrt{\sum_{j=1}^N\Delta^2
  x_j}/N=\Delta x/\sqrt{N}$, where $\Delta x_j$ is the spread of the
$j$th measurement (the spreads $\Delta x_j$s are all equal to $\Delta
x$: they refer to the same experiment). Notice that the same
$\sqrt{N}$ dependence can be obtained if, instead of using a classical
beam with $N$ average photons, we use $N$ separate single-photon
beams: in this case $\cos^2(\varphi/2)$ is the probability of the
photon exiting at output~D and $\sin^2(\varphi/2)$ is the probability
of the photon exiting at output~C.  The $1/\sqrt{N}$ bound on the
precision ($N$ being the number of photons employed) is referred to as
the shot noise limit. It is not fundamental and is only a consequence
of the employed classical detection strategy, where neither the state
preparation nor the readout takes advantage of quantum correlations.

Carefully designed quantum procedures can beat the $1/\sqrt{N}$ limit.
For example, injecting squeezed vacuum in the normally unused port~B
of the interferometer allows to achieve a sensitivity of
$1/N^{3/4}$~\cite{cavesprd,barnett}. Other strategies can do even
better, reaching an $1/N$ sensitivity with a $\sqrt{N}$ improvement
over the classical strategies detailed above. The simplest example
employs as the input to the interferometer the following entangled
state~\cite{dowlingatomgyro,yurke}
$$|\Psi\rangle=\frac
1{\sqrt{2}}\Big(|N_+\rangle_{\mbox{A}}|N_-\rangle_{\mbox{B}}+
|N_-\rangle_{\mbox{A}}|N_+\rangle_{\mbox{B}}\Big)\;,$$
where
$N_\pm\equiv(N\pm 1)/2$ and where the subscripts A and B label the
input ports. This is a highly non-classical signal where the
correlations between the inputs at~A and~B cannot be described by a
local statistical model. As before, the phase $\varphi$ can be
evaluated by measuring the photon number difference between the two
interferometer outputs, i.e.  by evaluating the expectation value of
the operator $M\equiv d^\dag d-c^\dag c=(a^\dag a-b^\dag
b)\cos\varphi+(a^\dag b+b^\dag a)\sin\varphi$, where $a$, $b$, $c$,
and $d$ are the annihilation operators of the optical modes at the
interferometer ports A, B, C, and D respectively (see
Fig.~\ref{f:mzint}).  This scheme allows a sensitivity of the order
$1/N$ for the measurements of small phase differences, i.e.
$\varphi\simeq 0$. In fact, the expectation value of the output photon
number difference is equal to $\langle M\rangle=-N_+\sin\varphi$ and
its variance is $\Delta^2M=\cos(2\varphi) + N_+^2 \sin^2\varphi$. The
error $\Delta\varphi$ on the estimated phase can be obtained from
error propagation, $\Delta\varphi=\Delta M/\left|\frac{\partial
    \langle M\rangle}{\partial\varphi}\right|$, and for $\varphi\simeq
0$, it is easy to see that it scales as $1/N$~\cite{dowlingatomgyro}.
Even though this procedure achieves good precision only for small
values of $\varphi$, other schemes exist that show the same high
sensitivity for all values of this parameter~\cite{sandersmilburn}.
Many quantum procedures that achieve the same $1/N$ sensitivity have
been proposed that do not make explicit use of entangled inputs.  For
example, one can inject into both interferometer inputs~A and~B
squeezed states and then measure the intensity difference at~C
and~D~\cite{jeffbondurant,kimbleshotnoise} or inject Fock states at~A
and~B, and then evaluate the photon-counting probability at the
output~\cite{burnett}, or, finally, measure the de Broglie wavelength
of the radiation~\cite{chuangdebroglie}.  One may wonder if this $1/N$
precision can be further increased, but in line with the time-energy
Heisenberg relation~\cite{mandelstamm} and the Margolus-Levitin
theorem~\cite{margolus}, it appears that this is a true quantum limit
and there is no way that it can be beaten~\cite{bollinger,ou}: it is
customarily referred to as the Heisenberg limit to interferometry.

\section*{\sf Quantum-enhanced parameter estimation}
\begin{figure}[hbt]
\begin{center}
\epsfxsize=.6\hsize\leavevmode\epsffile{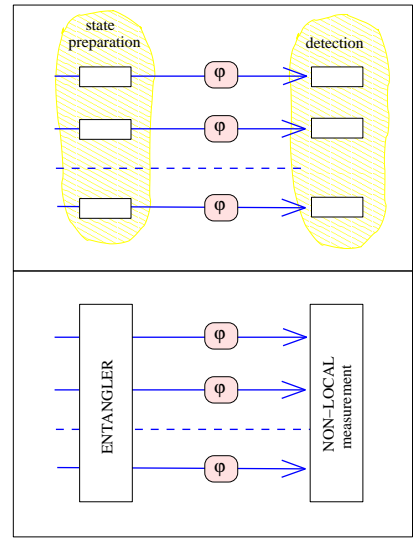}
\end{center}
\caption{Comparison between classical and quantum strategies. In
  conventional measurement schemes (upper panel), $N$ independent
  physical systems are separately prepared and separately detected.
  The final result comes from a statistical average of the $N$
  outcomes.  In quantum-enhanced measurement schemes (lower panel),
  the $N$ physical systems are typically prepared in a highly
  correlated configuration (i.e. an entangled or a squeezed state),
  and are measured collectively with a single non-local measurement
  that encompasses all the systems.} \label{f:qstrategy}\end{figure}

Some of the above interferometric techniques have found applications
also outside the context of optics, such as in
spectroscopy~\cite{bollinger} or in atomic
interferometry~\cite{jacobsonyamamoto}. In this section we point out a
general 
aspect of the quantum estimation theory on which most of the quantum
strategies presented in this review are based, i.e. the fact that
typically a highly correlated input is used and a collective
measurement is performed (see Fig.~\ref{f:qstrategy}). A simple
example~\cite{dowlingparameter} may help: consider a qubit, i.e. a
two-level quantum system which is described by the two states
$|0\rangle$ and $|1\rangle$ and their superpositions.  Suppose that
the dynamics leaves the state $|0\rangle$ unchanged and adds a phase
$\varphi$ to $|1\rangle$, i.e.  $|1\rangle\to e^{i\varphi}|1\rangle$.
If we want to estimate this phase, we can use a strategy analogous to
Ramsey interferometry by preparing the system in the quantum
superposition $|\psi_{in}\rangle\equiv(|0\rangle+|1\rangle)/\sqrt{2}$,
which is transformed by the system dynamics into
$|\psi_{out}\rangle=(|0\rangle+e^{i\varphi}|1\rangle)/\sqrt{2}$.  The
probability $p(\varphi)$ that the output state $|\psi_{out}\rangle$ is
equal to the input $|\psi_{in}\rangle$ allows to evaluate $\varphi$
as
$$p(\varphi)=\big|\langle\psi_{in}|\psi_{out}\rangle\big|^2=\cos^2(\varphi/2)\;.$$
This quantity can be estimated with a statistical error
$\Delta^2p(\varphi)=
\langle\psi_{out}|(|\psi_{in}\rangle\langle\psi_{in}|)^2|\psi_{out}\rangle-p^2(\varphi)=
p(\varphi)-p^2(\varphi)$. If we evaluate a parameter $\varphi$ from a
quantity $p(\varphi)$, error propagation theory tells us that the
error associated with the former is given by $\Delta\varphi=\Delta
p(\varphi)/\left|\frac{\partial p(\varphi)}{\partial\varphi}\right|$,
which in this case gives $\Delta \varphi=1$. We can improve such an
error by repeating the experiment $N$ times.  This introduces a factor
$1/\sqrt{N}$ in the standard deviation (again as an effect of the
central limit theorem) and we find an overall error
$\Delta\varphi=1/\sqrt{N}$.  (It is the same sensitivity achieved by
the experiment of a single photon in the interferometer described
above: these two procedures are essentially equivalent.) As in the
case of the interferometer, a more sensitive quantum strategy exists.
In fact, instead of employing $N$ times the state $|\psi_{in}\rangle$,
we can use the following entangled state that still uses $N$ qubits,
$$|\phi_{in}\rangle=\frac
1{\sqrt{2}}\Big(\underbrace{|0\rangle\cdots|0\rangle}_{N\mbox{
    times}}+\underbrace{|1\rangle\cdots|1\rangle}_{N\mbox{
    times}}\Big)$$
Now the tensor product structure of quantum
mechanics helps us, as the $e^{i\varphi}$ phase factors gained by the
$|1\rangle$s combine so that the corresponding output state is
$$|\phi_{out}\rangle=\frac 1{\sqrt{2}}\Big(|0\rangle\cdots|0\rangle+
e^{iN\varphi} |1\rangle\cdots|1\rangle\Big).$$
The probability
$q(\varphi)$ that $|\phi_{out}\rangle$ equals $|\phi_{in}\rangle$ is
$$q(\varphi)=\cos^2(N\varphi/2),$$
that, as before, can be estimated
with an error $\Delta^2q(\varphi)=q(\varphi)-q^2(\varphi)$.  This
means that $\varphi$ will have an error $\Delta\varphi=\Delta
q(\varphi)/\left|\frac{\partial
    q(\varphi)}{\partial\varphi}\right|=1/N$. This is a $\sqrt{N}$
enhancement over the precision of $N$ measurements on unentangled
qubits, which has been achieved by employing an entangled input and
performing a collective non-local measurement on the output, i.e. the
measurement of the probability $q(\varphi)$.

A generalization of the parameter estimation presented here is the
estimation of the input-output relations of an unknown quantum device.
A simple strategy would be to feed the device with a ``complete''
collection of independent states and measure the resulting outputs.
More efficiently, one can use entangled inputs: one half of the
entangled state is fed into the device and a collective measurement is
performed on the other half and on the device's
output~\cite{faith,dema}.  As in the case discussed above, the quantum
correlations between the components of the entangled state increase
the precision and hence reduce the number of measurements required. A
similar strategy permits to improve the precision in the estimation of
a parameter of an apparatus, or to increase the stability of
measurements~\cite{ntg}: part of an entangled state is fed into the
apparatus to be probed and an appropriate collective measurement is
performed on the output together with the other part of the entangled
state. This permits, for example, to discriminate among the four Pauli
unitary transformations applying the transformation only on a single
qubit probe. It would be impossible without entanglement.

\section*{\sf Quantum technology}
The quantum-enhanced parameter estimation presented above has found
applications in the most diverse fields.  In this section we give an
overview of some of them, leaving aside all the applications that
quantum mechanics has found in communication and
computation~\cite{chuang}, which are not directly connected with the
subject of this review.

{\bf Quantum frequency standards}~\cite{bollinger,fstand}. A typical
issue in metrology and spectroscopy is to measure time or frequency
with very high accuracy.  This requires a very precise clock, i.e. an
oscillator.  Atomic transitions are so useful to this aim that the
very definition of second is based on them. To measure time or
frequency accurately, we can start with $N$ cold ions in the ground
state $|0\rangle$ and apply an electromagnetic pulse that creates
independently in each ion an equally weighted superposition
$(|0\rangle+|1\rangle)/\sqrt{2}$ of the ground and of an excited state
$|1\rangle$.  A subsequent free evolution of the ions for a time $t$
introduces a phase factor between the two states that can be measured
at the end of the interval by applying a second, identical,
electromagnetic pulse and measuring the probability that the final
state is $|0\rangle$ (Ramsey interferometry). This procedure is just a
physical implementation of the qubit example described above, but here
the phase factor is time dependent and is equal to $\varphi=\omega t$,
where $\omega$ is the frequency of the transition
$|0\rangle\leftrightarrow|1\rangle$.  Hence the same analysis applies:
from the $N$ independent ions we can recover the pursued frequency
$\omega$ (from the phase factor $\varphi$) with an error
$\Delta\varphi=1/\sqrt{N}$, i.e.  $\Delta\omega=1/(\sqrt{N}t)$.

Instead of acting independently on each ion, one can start from the
entangled state $|\phi_{in}\rangle$ introduced above. In this case,
the error in the determination of the frequency is
$\Delta\omega=1/(N t)$, i.e. there is an enhancement of the
square root of the number $N$ of entangled ions over the previous
strategy.  

{\bf Quantum lithography and two-photon
  microscopy}~\cite{scullylith,salehmicr,dowl,shihlith,ganna}.  When
we try to resolve objects smaller than the wavelength of the employed
light, the wave nature of radiation becomes important: the light tends
to scatter around the object, limiting the achievable resolution. This
defines the Rayleigh diffraction bound, which restricts many optical
techniques: it is not always practical to reduce the wavelength.
Quantum effects can help by decreasing the wavelength of the light
while keeping the wavelength of the radiation field constant. How can
this, apparently paradoxical, effect come about? The basic idea is to
employ physical devices that are sensitive to the de Broglie
wavelength: in quantum mechanics to every object we can associate
wavelength $\lambda=2\pi\hbar/p$ where $p$ is the object's momentum
(for radiation $p$ is the energy $E$ divided by the speed of light
$c$). Obviously the wavelength of a single photon $\lambda=2\pi\hbar
c/E=2\pi c/\omega$ is the wavelength of its radiation field. But what
happens if we are able to employ a ``biphoton'' (i.e. a single entity
constituted by two photons)? In this case we find that its wavelength
is $2\pi\hbar c/(2E)=\lambda/2$: half the wavelength of a single
photon, or equivalently half the wavelength of its radiation field. Of
course, using ``triphotons'', ``quadriphotons'', {\em etc.}  would
result in further decreases of wavelengths. Experimentalists are able
to measure the de Broglie wavelengths of
biphotons~\cite{monken,shihlith,chuangdebroglie}, so that
theoreticians have concocted useful ways to employ them: the most
important applications are quantum lithography~\cite{scullylith,dowl},
where smaller wavelengths help to etch smaller integrated circuit
elements on a two-photon sensitive substrate, and two-photon
microscopy~\cite{ganna}, where they produce less damage to the
speciments. Also in this context, entanglement is a useful resource as
it is instrumental in creating the required biphotons and in enhancing
the cross-section of two-photon absorption~\cite{salehmicr}.

{\bf Quantum positioning and clock
  synchronization}~\cite{nature,conveyorbelt,jozsa}.  To find out the
position of an object, one can measure the time it takes for some
light signals to travel from that object to some known reference
points.  The best classical strategy is to measure the travel times of
the single photons in the beam and to calculate their average. This
allows to determine the travel time with an error proportional to
$1/(\Delta\omega \sqrt{N})$, where $\Delta\omega$ is the signal
bandwidth which induces a minimum time-duration of $1/\Delta\omega$
for each photon, i.e. the times of arrival of each of the photons will
have a spread $1/\Delta\omega$. The accuracy of the travel time
measure thus depends on the spectral distribution of the employed
signal.  The reader will bet that a quantum strategy allows to do
better with the same resources. In fact, by entangling $N$ photons in
frequency, we can create a ``super-photon'' whose bandwidth is still
$\Delta\omega$ (i.e. it employs the same energetic resources as the
$N$ photon signal employed above), but whose mean effective frequency
is $N$ times higher, as the entanglement causes the $N$ photons to have
the same frequency. This means that the super-photon allows us to
achieve $N$ times the accuracy of a single photon with the same
bandwidth. To be fair, we need to compare the performance of the
super-photon with that of a classical signal of $N$ photons, so that
the overall gain of the quantum strategy is $\sqrt{N}$~\cite{nature}.

The problem of localization is intimately connected with the problem
of synchronizing distant clocks. In fact, by measuring the time it
takes for a signal to travel to known locations, it is possible to
synchronize clocks at these locations. This immediately tells us that
the above quantum protocol can give a quantum improvement in the
precision of distant clocks synchronization. Moreover, quantum effects
can be also useful in avoiding the detrimental effects of
dispersion~\cite{franson}.  The speed of light in dispersive media has
a frequency dependence, so that narrow signals (which are constituted
by many frequencies) tend to spread out during their travel. This
effect ruins the sharp timing signals transmitted. Using the non-local
correlations of entangled signals, we can engineer frequency-entangled
pulses that are not affected by dispersion and that allow clock
synchronization~\cite{conveyorbelt}.

{\bf Quantum imaging}~\cite{kolobov,lugiato}. A large number of
applications based on the use of quantum effects in spatially
multimode light can be grouped under the common label of quantum
imaging. 

The most famous quantum imaging experiment is the reconstruction of
the so called ``ghost images''~\cite{ghosts}, where non-local
correlations between spatially entangled two-photon states are used to
create the image of an object without directly looking at it. The
basic idea is to illuminate an object with one of the twin photons
which is then absorbed by a ``bucket'' detector (it has no spatial
resolution but is just able to tell whether the photon crossed the
object or whether it was absorbed by it). The other entangled photon
is shone onto an imaging array and the procedure is repeated many
times.  Correlating the image on the array with the coincidences
between the arrival of one photon at the bucket detector and the other
at the imaging array, the shape of the object can be determined. This
is equivalent to the following scenario: use a device that shoots two
pebbles in random but exactly opposite directions.  When one of the
pebbles overshoots an object, it hits a bell. The other pebble instead
hits a soft wall, where it remains glued to it. By shooting many
pebbles and marking on the wall the pebble's position every time we
hear the bell ring, we will project the outline of the object on the
wall. The fact that such an intuitive description exists should make
us queasy on the real quantum nature of the ghost image experiment
and, in fact, it was shown that even if it makes use of highly
non-classical states of light, it is an essentially classical
procedure~\cite{boyd}. The quantum nature of such an experiment lies
in the fact that, using the same apparatus, both the near-field and
the far-field plane can be perfectly imaged.  Classical correlations
do not allow this, even though classical thermal light can approximate
it~\cite{lugiato2}. A related subject is the creation of noiseless
images or the noiseless image amplification~\cite{lugiato,kolobov},
i.e. the formation of optical images whose amplitude fluctuations are
reduced below the shot-noise and can be, in principle, suppressed
completely.

Many applications require us to measure very accurately the direction
in which a focused beam of light is shining: a typical example is
atom-force microscopy, where the deflection of a light beam reflected
from a cantilever that feels the atomic force can achieve nanometric
resolution. As a light beam is, ultimately, composed of photons, the
best way to measure its direction is apparently to shine the beam on
an infinitely resolving detector, to measure where each of the photon
is inside the beam and to take the average of the positions.  This
strategy will estimate the position of the beam with an accuracy that
scales as $\Delta d/\sqrt{N}$, where $\Delta d$ is the beam width and
$N$ is the number of detected photons. Analogously as for the
shot-noise, this limit derives from the quantized nature of light and
from the statistical distribution of the photons inside the beam. As
in interferometry, also here quantum effects can boost the sensitivity
up to $1/N$~\cite{fabrebeam,barnett,treps}. In fact, consider the
following simple example, in which the beam shines along the $z$
direction and is deflected only along the $x$ direction.  We can
measure such a deflection by shining it exactly between two perfectly
adjacent detectors and measuring the photon number difference between
them.  If we expand the spatial modes of the light beam into the sum
of an ``even'' mode which is symmetrical in the $x$ direction and an
``odd'' mode which changes sign at $x=0$ (between the two detectors),
we see that the beam is perfectly centered when only the even mode is
populated and the odd mode is in the vacuum.  Borrowing from sub-shot
noise interferometry, we see that we can achieve a $1/N^{3/4}$
sensitivity by populating the odd mode with squeezed vacuum instead.
Moreover, we can achieve the Heisenberg limit of $1/N$ by populating
both modes with a Fock state $|N/2\rangle$~\cite{barnett}.

Any object that creates an image (e.g. a microscope or a telescope) is
necessarily limited by diffraction, due to its finite transverse
dimensions. Even though classical ``super-resolution'' techniques are
known that can be used to beat the Rayleigh diffraction limit, these
are ultimately limited by the quantum fluctuations that introduce
undesired quantum noise in the reconstructed image. By illuminating
the object with bright multimode squeezed light and by replacing the
part which the finite dimensions of the device cuts away with squeezed
vacuum, we can increase the resolution of the reconstructed
image~\cite{fabre}, at least in the case of weakly absorbing objects
(opaque objects would degrade the squeezed light shining on them).

{\bf Coordinate transfer}~\cite{popescu,perescoord,barcellona,giulio}.
A peculiar example of quantum enhanced strategy arises in the context
of communicating a direction in space~\cite{popescu} or a reference
frame~\cite{perescoord,barcellona,giulio} (composed by three
orthogonal directions). If there is no prior shared reference, it
requires some sort of parallel transport such as exchanging gyroscopes
(which in quantum mechanical jargon are called spins).  Quantum
mechanics imposes a bound on the precision with which the axis of a
gyroscope can be measured, as the different components of the
angular momentum are incompatible observables: unless one knows the
rotation axis {a priori}, it is impossible to measure exactly the
total angular momentum. Gisin and Popescu found the baffling result
that sending two gyroscopes pointing in the same direction is less
efficient (i.e. allows a less accurate determination of this
direction) than sending two gyroscopes pointing in opposite
directions~\cite{popescu}.  The reason is that the most efficient
measurement to recover an unknown direction from a couple of spins is
an entangled measurement, i.e. a measurement that has operators with
entangled eigenvectors associated to it. Such a detection strategy
cannot be separated in different stages, so that it is not possible to
rotate the apparatus before the measurement on the second spin which
would imply the equivalence of the two scenarios. The two scenarios
could be shown equivalent also if it were possible to flip the
direction of the second spin without knowing its rotation axis, but
this is impossible (it is an anti-unitary transformation whereas
quantum mechanics is notably unitary). Elaborating on this idea, many
quantum enhanced coordinate transfer
strategies~\cite{perescoord,barcellona,giulio} have been found.

\section*{\sf Repeated position measurements: beating the standard quantum
  limit} The continuous measure of the position of a free mass is a
paradigmatic example of how classical strategies are limited in
precision. This experiment is typical of gravitational-wave detection,
where the position of a test mass must be accurately monitored. The
standard quantum limit~\cite{sql,edelsteinjpe,rmp,brajinskij} arises
in this context by directly applying the Heisenberg relation to two
consecutive measurements of the position of the free mass, without
taking into account the possibility that the first measurement can be
tuned to change appropriately the position configuration of the mass.
The original argument was the following: suppose that we perform the
first position measurement at time $t=0$ with an uncertainty $\Delta
x(0)$. This corresponds (via the Heisenberg uncertainty relation
---see Fig.~\ref{f:heisenb}) to an uncertainty in the initial momentum
$p$ at least equal to $\Delta p(0)=\hbar/[2\Delta x(0)]$. The dynamics
of an unperturbed the free mass $m$ is governed by the Hamiltonian
$H=p^2/2m$, which evolves at time $t$ the position as
$x(t)=x(0)+p(0)t/m$.  This implies that the uncertainty in the initial
momentum $p(0)$ transfers into an uncertainty in the position $x(t)$.
The net effect appears to be that a small initial uncertainty
$\Delta^2 x(0)$ produces a big final uncertainty $\Delta^2
x(t)\simeq\Delta^2x(0)+\Delta^2p(0)\;t^2/m^2\geq 2\Delta x(0)\Delta
p(0)\;t/m\geq\hbar t/m$. In this derivation there is an implicit
assumption that the final uncertainty $\Delta x(t)$ cannot be
decreased by the correlations between the position and the momentum
that build up during the unitary evolution after the first
measurement. This is unwarranted: Yuen showed that an exotic detection
strategy exists which, after the first measurement, leaves the mass in
a ``contractive state''~\cite{yuen}, i.e.  whose position uncertainty
decreases for a certain period of time. [The time $t$ for which a mass
in such a state has a spread in position $\Delta^2 x(t)$ below a level
$2\delta^2\hbar/m$ satisfies $t\leq 4\delta^2$].  The standard quantum
limit is beaten, i.e. $\Delta^2 x(t)\leq\hbar t/m$, if the second
measurement is performed soon enough.  The debate evolved then to
ascertaining wether two successive measurements at times $0$ and $t$
can be performed both of which beat the standard quantum
limit~\cite{cavesprl}. In fact, a simple application of the Heisenberg
relation gives $\Delta x(0)\Delta x(t)\geq\frac
12\left|\left\langle\left[x(0),x(t)\right]\right\rangle\right|=\hbar
t/2m$, from which it seems impossible that both measurements have a
spread $\leq\sqrt{\hbar t/2m}$. However, $\Delta x(0)$ is the variance
of the state immediately after the first measurement, which does not
necessarily coincide with the variance of the results of the first
measurement. In fact, it is possible~\cite{ozawa} to measure the
position accurately and still leave the mass in a contractive state
with initial variance $\Delta x(0)\gg\sqrt{\hbar t/2m}$, so that the
standard quantum limit can be beaten also repeatedly.

Notice that the back-action introduced in the derivation of the
standard quantum limit would not occur if one were to measure the
momentum instead of the position, since the above Hamiltonian
conserves the momentum, $p(t)=p(0)$, which is independent of the
position. The momentum measure is an example of a quantum
non-demolition detection
scheme~\cite{rmp,grangierqnd,brajinskij,wallqnd}, in which one removes
any feed-back in the detection by focusing on those observables which
are not coupled by the dynamics to their incompatible counterparts.

The standard quantum limit arises also in the context of
interferometric measurements of
position~\cite{edelsteinjpe,cavesprl1,brajinskij}, where the mass is
typically one of the mirrors of the interferometer.  The movement of
the mirror introduces a phase difference between the arms of the
interferometer (see Fig.~\ref{f:mzint}). To achieve high measurement
precision, one is hence tempted to feed the interferometer with
electromagnetic signals that posses a well defined phase.  However,
the phase and the intensity of the electromagnetic field are in some
sense complementary and a well defined phase corresponds to a highly
undetermined intensity. At first sight this seems without
consequences, but any mirror feels a force dependent on the intensity
of the light shining on it, through the mechanism of radiation
pressure. Hence, the fluctuations in intensity of a signal with
well-defined phase induce a fluctuating random force on the mirror
which ultimately spoils the precision of the measurement setup. Using
sufficiently intense coherent light and optimizing the phase and
intensity fluctuations, one finds that the attainable precision is
again the standard quantum limit~\cite{edelsteinjpe,cavesprl1}.
Apparently, this derivation of the standard quantum limit is
completely independent from the one given above starting from the
Heisenberg relation. However, also here there is an unwarranted
assumption, i.e. the treatment of phase and intensity fluctuations as
independent quantities.  Caves showed that, by dropping this premise,
one can do better~\cite{cavesprd}.  In fact, a squeezed input signal
(see Fig.~\ref{f:heisenb}) where the amplitude quadrature has less
quantum fluctuations than the phase quadrature produces a reduced
radiation pressure noise at the expense of an increased
photon-counting noise, and vice-versa.  This balance allows to
fine-tune the parameters so that the standard quantum limit can be
reached with much lower light intensity. Refinements of this technique
allows to even beat the standard quantum limit by tailoring
appropriate squeezed
states~\cite{unruh,jeffbondurant,reynaud,walls93,brajinskij} or by
employing quantum non-demolition measurements~\cite{rmp}.

The standard quantum limit is not a fundamental precision threshold.
However, at present its conquest is still an open experimental
challenge.  In fact, on one hand, most of the above theoretical
proposals are quite impractical and should be seen only as proofs of
principle and, on the other hand, many competing sources of noise
become important when performing very precise measures. The most
important is, of course, the thermal fluctuations in the mass to be
monitored, but the shot noise at the detection stage or the
dissipative part of the mirror response are also big
limitations~\cite{rmp,onofrio,naturelevers,sciencelevers}.  Various
techniques to beat this threshold have been proposed.  Among others
(by necessity the following list is incomplete), we can cite the
techniques to employ feedback techniques to enforce a positive
back-action~\cite{wiseman,stefano,cohadon,darianofeedback}, or the
huge number of techniques to perform quantum non-demolition
measurements~\cite{rmp,imoto,grangierqnd,haroche,grangiernat}, or to
build contractive states, or to build speed-meters~\cite{speed}.  At
the present stage, the most promising seem the use of
nanotechnologies, where tiny mechanical oscillators are coupled to
high sensitivity electronics~\cite{naturelevers,sciencelevers}, or the
new generations of gravitational wave detectors~\cite{ligo2}.

\section*{\sf Quantum limits to the measurement of spacetime geometry} 
Quantum effects can be used to increase the accuracy of many different
kinds of measurements, but which are the ultimate limits to the
resolution that the physical laws allow?  Attempts to derive quantum
limits to the accuracy of measuring the geometry of spacetime date
back at least to Wigner~\cite{wigner,witten}.  As the preceding discussions
show, however, care must be taken in applying non-fundamental bounds
such as the standard quantum limit.  Fortunately, the
Margolus-Levitin~\cite{margolus} theorem and techniques from the
physics of computation~\cite{sethnature,sethprl} can be used to derive
limits to the accuracy with which quantum systems can be used to
measure spacetime geometry.

The first question is that of minimum distance and time.  One can
increase the precision of clocks used to measure time by increasing
their energy: the Margolus-Levitin theorem implies that the minimum
`tick length' of a clock with energy $E$ is $\Delta t = \pi \hbar/2E$.
Similarly, the wavelength of the particles used to map out space can
be decreased by increasing their energy.  There appears to be no
fundamental physical limit to increasing the energy of the clocks used
to measure time and the particles used to measure space, until one
reaches the Planck scale, $t_P = \sqrt{\hbar G/ c^5} = 5.391 \times
10^{-44}$ seconds, $\ell_P = ct_P$.  At this scale, the Compton
wavelength $2\pi\hbar/mc$ of the clocks and particles is on the same
order of magnitude as their Schwarzschild radius $2mG/c^2$, and
quantum gravitational effects come into play~\cite{qgravity}.  The
second question is that of the accuracy to which one can map out the
large-scale structure of spacetime.  One way to measure the geometry
of spacetime is to fill space with a `swarm' of clocks, exchanging
signals with the other clocks and measuring the signals' times of
arrival.  In this picture, the clocks could be as large as GPS
satellites, or as small as elementary particles.

Let's look at how accurately this swarm of clocks can map out a volume
of spacetime with radius $R$ over time $T$.  Every tick of a clock or
click of a detector is an elementary event in which a system goes from
a state to orthogonal state.  Accordingly, the total number of ticks
and clicks that can take place within the volume is a scalar quantity
limited by the Margolus-Levitin theorem: it is less than
$2ET/\pi\hbar$, where $E$ is the energy of the clocks within the
volume.

If we pack the clocks too densely, they will form a black hole and be
useless for the measurement of spacetime outside their horizon.  To
prevent black hole formation, the energy of clocks within a spacelike
region of radius $R$ must be less than $Rc^4/2G$.  As a result, the
total number of elementary events that can occur in the volume of
spacetime is no greater than
$$N \equiv \pi^{-1}(T/t_P)(R/l_P). \eqno[1]$$
This quantum geometric
limit can also be formulated in a covariant fashion. The maximum
number of ticks and clicks in a volume is a scalar quantity
proportional to the integral of the trace of the energy-momentum
tensor over the four volume; and $2TR$ can be identified with the area
of an extremal world sheet contained within the four volume.  The
quantum geometric limit of Eq.~[1] was derived without any recourse to
quantum gravity: the Planck scale makes its appearance simply from
combining quantum limits to measurement with the requirement that a
region not itself be a black hole.  (If the region is at or above its
critical density, then Eq.~[1] still holds if $R$ is the radius of the
horizon of the region as measured by an external observer.)

The quantum geometric limit is consistent with and complementary to
the Bekenstein bound, the holographic bound, and the covariant entropy
bound~\cite{bekenstein,thooft,susskind,bousso}, all of which limit the
number of bits that can be contained within a region.  [It also
confirms Ng's prediction~\cite{ng} for the scale of spacetime foam.]
For example, the argument that leads to Eq.~[1] also implies that
maximum number of quanta of wavelength $\lambda \leq 2R$ that can be
packed into a volume of radius $R$ without turning that volume into a
black hole is bounded by $R^2/\pi l_P^2$~\cite{yurts}, in accordance
with the Bekenstein bound and holography.  Because it bounds the
number of elementary events or `ops,' rather than the number of bits,
the quantum geometric limit~[1] implies a tradeoff between the
accuracy with which one can measure time, and the accuracy with which
one can measure space: the maximum spatial resolution can only be
obtained by relaxing the temporal resolution and having each clock
tick only once in time $T$.  This lack of temporal resolution is
characteristic of systems, like black holes, that attain the
holographic bound~\cite{sethnature}.  By contrast, if the events are
spread out uniformly in space and time, the number of cells within the
spatial volume goes as $(R/l_P)^{3/2}$-- less than the holographic
bound-- and the number of ticks of each clock over time T goes as
$(T/t_P)^{1/2}$.  This is the accuracy to which ordinary matter such
as radiation and massive particles map out spacetime.  Because it is
at or close to its critical density, our own universe maps out the
geometry of spacetime to an accuracy approaching the absolute limit
given by $R^2/\pi l_P^2$: there have been no more than $(T/t_P)^2
\approx 10^{123}$ `ticks and clicks' since the big
bang~\cite{sethprl}.

\section*{\sf Conclusion}
Quantum mechanics governs every aspect of the physical world,
including the measuring devices we use to obtain information about
that world.  Quantum mechanics limits the accuracy of such devices via
the Heisenberg uncertainty principle and the Margolus-Levitin theorem;
but it also supplies quantum strategies for surpassing semi-classical
limits such as the standard quantum limit and the shot noise limit.
Starting from strategies to enhance the sensitivity of interferometers
and position measurements, scientists and engineers have developed
quantum technologies that employ effects such as squeezing and
entanglement to improve the accuracy of a wide variety of
measurements.  Some of these quantum techniques are still futuristic:
at present, methods for creating and manipulating entangled states are
still in their infancy.  As we saw, quantum effects usually allow a
precision enhancement equal to the square root of the number $N$ of
employed particles; but it is usually very complicated to entangle as
few as $N = 5$ or $6$ particles.  In contrast, it is typically rather
simple to employ millions of particles to use the classical strategy
of plain averaging.  As quantum technologies improve, however, the use
of entanglement and squeezing to enhance precision measurements is
likely to become more wide spread.  Meanwhile, as the example of
quantum limits to measuring spacetime geometry shows, examining the
quantum limits to measurement can give insight into the workings of
the universe at its most fundamental levels.

{\bf{\textsf{Acknowledgements:}}} VG acknowledges financial
support by EC contracts IST-SQUIBIT, IST-SQUBIT2, and RTN-Nanoscale
Dynamics. SL was supported by DARPA, ARDA and ARO via a MURI program.
LM acknowledges financial support by EC ATESIT project IST-2000-29681
and MIUR Cofinanziamento 2003.
\end{document}